\newif\ifjournalstyle	
\journalstylefalse	

\newif\iffigs
\figsfalse		

\ifjournalstyle				
	\documentstyle{l-aa}
	\def\emph#1{{\em #1\/}}
	\def\mathrm#1{{\rm #1}}
	\figsfalse			
\else
	\makeatletter
	\@ifundefined{documentclass}{	
		\documentstyle{article}
		\def\emph#1{{\em #1\/}}	
		\@ifundefined{mathrm}{\def\mathrm#1{\rm #1}}{}	
		\figsfalse
		}{
		\documentclass[a4paper,twocolumn]{article}	
		\iffigs\usepackage{graphics}\fi
		}
	\makeatother
\fi

\title{Prominence and flare fine structure from cross-field
	thermal conduction\ifjournalstyle
		\else\thanks{To appear in {\it Astronomy and Astrophysics}.
		LANL preprint archive number astro-ph/9509011}\fi}
\ifjournalstyle
	\author{Norman Gray\inst{1}\and John C Brown\inst{1}}
	\institute{Dept.\ of Physics and Astronomy, University of Glasgow,
		Glasgow G12 8QQ, U.K.}
	\offprints{Norman Gray}
	\date{Received June 1995; accepted 29 August 1995}
	\thesaurus{9(02.16.1; 06.06.3; 06.06.2)}
\else
	\author{Norman Gray\thanks{{\tt norman@astro.gla.ac.uk}}
		and John C Brown\\
		{Dept.\ of Physics and Astronomy, University of Glasgow,
		Glasgow G12 8QQ, U.K.}}
	\date{29 August 1995}
\fi

\makeatletter
\let\@internalcite\cite
\def\cite{\@ifstar{\citeyear}{\citefull}}
\def\citefull{\def\astroncite##1##2{##1, ##2}\@internalcite}
\def\citeyear{\def\astroncite##1##2{##2}\@internalcite}

\def\@citex[#1]#2{\if@filesw\immediate\write\@auxout{\string\citation{#2}}\fi
  \def\@citea{}\@cite{\@for\@citeb:=#2\do
    {\@citea\def\@citea{; }\@ifundefined
       {b@\@citeb}{{\bf ?}\@warning
       {Citation `\@citeb' on page \thepage \space undefined}}%
{\csname b@\@citeb\endcsname}}}{#1}}

\def\@cite#1#2{(#1\if@tempswa , #2\fi)}
\def\@biblabel#1{}
\makeatother

\newcommand{\be}[1]{\begin{equation}\label{e:#1}}
\newcommand{\ee}{\end{equation}}
\newcommand{\bea}{\begin{eqnarray}}
\newcommand{\eea}{\end{eqnarray}}
\newcommand{\bean}{\begin{eqnarray*}}
\newcommand{\eean}{\end{eqnarray*}}
\newcommand{\heat}{L}
\newcommand{\thickness}{l}
\newcommand{\pa}{\|}
\newcommand{\pe}{\bot}
\newcommand{\Btau}{\tilde\tau}	

\makeatletter

\def\displaylineseqno#1{\refstepcounter{equation}\label{e:#1}%
	\hbox{\@eqnnum}\cr}

\let\draftnote\@gobble

\def\units{\@ifstar{\let\un@tsspace\relax    \un@ts}%
                   {\let\un@tsspace\thinspace\un@ts}}
\newcommand{\un@ts}[1]{{\let~\,\ifmmode \un@tsspace\mathrm{#1}\else
	\nobreak$\un@tsspace\mathrm{#1}$\fi}}

\makeatother

\def\:#1#2{{\scriptstyle {#1\over#2}}}	
\newcommand{\dd}{{\mathrm d}}		
\def\ddd#1#2{{\dd #1\over\dd #2}}
\newcommand{\eqnref}[1]{Eqn.~(\ref{e:#1})}
\newcommand{\figref}[1]{Fig.~\ref{f:#1}}

\ifjournalstyle\else
\newenvironment{acknowledgements}{\emph{Acknowledgements:}}{}

\newcommand{\la}{\mathrel{\vcenter{\hbox{\ooalign{\raise 4.75pt
	\hbox{$<$}\crcr $\sim$}}}}}
\newcommand{\ga}{\mathrel{\vcenter{\hbox{\ooalign{\raise 4.75pt
	\hbox{$>$}\crcr $\sim$}}}}}
\fi

\begin{document}
\maketitle

\begin{abstract}
Thermal conduction across a mag\-net\-ic field is  \linebreak[4]
strong\-ly sup\-pressed
com\-pared with conduction along the field.  However, if a flare is
heated by a highly filamented beam directed along the field, then the
array of heated cells in
a cross-section of the flare will result in both small spatial scales
(with consequently large temperature gradients) and a large surface
area for the heated volume, providing a geometrical enhancement of the
total cross-field energy flux.  To investigate the importance of this
filamentary geometry, we present a simple model of a single heated
filament surrounded by an optically thin radiating shell, obtain an
analytical expression for the stable equilibrium temperature profile
within the shell, and use this to impose limits on the size of
filament for which this model is appropriate.

We find that this mechanism by itself is capable of transporting a
power of the same order as a large flare, with a moderate range of
filament sizes.  The length scales are substantially smaller
than can be resolved at present, although they should be regarded as
underestimates.

\ifjournalstyle%
\else {\bf Key words:} Plasmas -- Sun: flares -- Sun: filaments
\fi
\end{abstract}

\section{Introduction}

There is an observational
consensus \cite{engvold76,chiuderidrago92} that solar prominences have
a good deal of fine structure within them, on scales ranging {from}
100~km to 1500~km; similarly, each improvement in
resolution\draftnote{`Standard' Yohkoh and NIXT references\dots} finds more
structure
within flares.  Other studies, such as Fletcher and
Brown~\cite*{fletcher95}, provide indirect support for the existence
of fine filamentary structure in flares, by invoking the presence of
such structure to explain polarisation observations.  At the same
time, there seems to be a theoretical consensus both that prominences
condense due to a thermal instability~\cite{field65} in the plasma
around them, and that cross-field thermal conduction plays a
significant role in determining the scale of their fine
structure~\cite{vanderlinden93}.  However, these studies (as reviewed
by~\cite{heyvaerts74}) concentrate
on the dynamics of the instability which forms the threads, rather
than on the final equilibrium of the resulting structure.  In this
paper, we instead present a simple, quasi-static, analytical model of a
filamentary structure in a prominence or flare, which exhibits the
small spatial scales necessary to produce a temperature gradient large
enough to make cross-field thermal conduction significant, and which
potentially gives cross-field conduction a dominant role in governing
both the number and size of the threads in a prominence or flare, and
the energy budget of the beam or current supplying the thread.

As discussed in a preliminary version of this model~\cite{brown94}, we
have available a large number of free parameters, so the
aim has to be not to make a single prediction, but rather to see which
combinations of parameters allow a consistent model.

The paper is organised as follows: in section~\ref{s:model} we
describe the model we have used; in section~\ref{s:profiles} we
describe the analytical and numerical solutions to this model that we
have obtained; in section~\ref{s:stability} we discuss the stability
of these solutions; and in section~\ref{s:lowbeta} we briefly discuss
the minor changes to our results which are necessary in the case of a
low-$\beta$ plasma.

\section{The model}
\label{s:model}

We suppose the existence of a heating mechanism which is spatially
fragmented with filaments aligned along the ambient magnetic field, and so
divides the heated plasma into a number of identical cylindrical
cells, radius~$R$, each of which is heated only within some coaxial
cylindrical core (we have also considered the case of filamentation
into plane sheets, with similar results to those presented here).
This heated core is surrounded by a shell which is
heated solely by conduction {from} the core, and cooled solely by
radiation.  We find the temperature structure of this shell which
allows an equilibrium to form by radiating away all the energy that is
transported into it.  We can ignore the effects of longitudinal
conduction, and assume that the filamentation allows the transverse
conduction to dominate -- this is retrospectively justified in
Sect.~\ref{s:comparison}.
Thus, the equilibrium structure of a particular `slice' (of
length~$\thickness$) through the cell will be governed only by the
power input at a particular point along the cylinder.  When
we wish to find numerical values, below, we will take
$\thickness=10^7\units m$ as the cylinder length (this is the
approximate stopping distance for a typical electron beam).

We have been deliberately vague about the heating mechanism -- we will
find that the only relevant parameter is the total power per unit
length conducted {from} within the core, so that this heating mechanism could
equally well be a non-thermal beam or a dissipating current.  The cost
of this generality is that we can say nothing about the structure
of the core, nor fix what fraction of the total power available is
conducted out, rather than being carried longitudinally within the
core.  We expect, however, that the heat-loss {from} the core will be
substantial.

We parameterise distances {from} the centre of the cell in terms of
the dimensionless parameter $s=r/R$, where $s=1$ at the edge of the
cell.  We will take the shell to consist of material below some
temperature~$T_\rho$ which we will take, below, to be around
the peak in the Cox-Tucker radiative loss function.  This shell
starts a distance~$\rho R$ {from} the centre of the cell,
so that~$\rho$ is defined by
\be{Trho}
	T(s=\rho)=T_\rho.
\ee
Conduction need not be the only method by which the core is cooled,
but we will denote by~$P_\rho$ the power actually conducted into the
shell per unit length along the field.  Finally, we take each cell to
be independent, so that there is no energy flux between cells; thus
the temperature gradient must vanish at $s=1$.

We emphasise that all the power input is confined within the
radius~$\rho$, and that we will not concern ourselves further with the
details of this input nor the internal structure of the core.  For
this to be valid we must, for example, take the plasma in the shell to
be optically thin so that it will not experience significant secondary
heating by any radiative emission {from} the core.

In the shell, there will be two processes competing: there will be a
conductive heat flux entering {from} the core, and radiative emission
{from} the
heated plasma.  For~$\rho\leq s\leq1$, the conductive heat flux per
unit axial length is
\be{P12}
	P(s) = Q(s) 2\pi s R,
\ee
where the heat flux density~$Q(s)$ is (neglecting $\kappa_\pa \dd
T/\dd r$ along the beam)
\be{Q12}
	Q(s) = -\kappa_\pe(s) {\dd\over\dd r} T(s),
\ee
and the cross-field thermal conduction coefficient
is \cite{rosenbluth58}
\be{kpdef}
	\kappa_\pe = \kappa_1 {n^2\over B^2 T^{1/2}}
\ee
with~$n$ in~\units{m^{-3}}, $B$ in Tesla, and
\[
	\kappa_1 \equiv 9.76\times10^{-41}\units{W~m^5~T^2~K^{-1/2}}
\]
(note that the effective coefficient is likely to be rather larger
than this -- see the discussion in section~\ref{s:turb} below).
High in the solar atmosphere, magnetic pressure dominates, and we take
the magnetic field to be approximately the
same across the filament, so that the pressure-balance
condition reduces to
\be{pbalance}
	n(s) T(s) = a,\quad\hbox{(constant)}
\ee
(but see Sect.~\ref{s:lowbeta}) and in terms
of~$\kappa_0\equiv\kappa_1/B^2$ we can write
\be{kp}
	\kappa_\pe = \kappa_0 (B) {n^2\over T^{1/2}}
		= \kappa_0 (B) a^2 T^{-5/2}.
\ee

Given that the emitted power {from} a vol\-ume $\dd V$ is \linebreak[3]
$n^2f(T) \dd V$, for a given an\-nulus at radius $s$, we will have
\[
	P(s) - P(s+\dd s) - n(s)^2 f(T(s)) \cdot 2\pi sR \cdot R\dd s
		= 0,
\]
which, on substitution of \eqnref{P12}, reduces to
\be{Qtaueq}
	{\dd\over\dd s}[sQ(s)] + n^2 f(T) s R = 0.
\ee
The vanishing of the radial flux at the outer boundary of the
slice provides one boundary condition $Q(1)=0$, and another will be
provided indirectly by the specified total power input to the shell at
radius~$\rho$.

Before going on to determine the temperature profile explicitly, it is
worthwhile to point out that the forms of the transverse conduction
coefficient (\eqnref{kp}) and radiative loss function~$f(T)$ conspire
to produce a temperature profile which is radically different {from}
the profile
produced by longitudinal conduction (as in the usual treatment of
flares and the transition region).  Because~$\kappa_\pe$ depends on
a substantial negative power of the temperature, the thermal
conductivity \emph{improves} for lower temperature, in the sense that a
shallower gradient is needed to carry a given heat flux.  Between
$T=10^4\units K$ and $10^5\units K$ the radiative loss
function~$f(T)\sim T^2$ and this decrease in the efficiency of
radiative loss at lower temperature, along with the improvement
in~$\kappa_\pe$ there, means that the temperature gradient is very
shallow at the cold outer edge of the shell, and steepens particularly
sharply at the heated inner boundary, quite unlike the
profile in a normal gas or solid.

\section{Analytical temperature profiles}
\label{s:profiles}

The function~$f(T)$ has an intricate dependence on temperature, but
for our purposes we may take it to be a series of power laws
\cite{rosner78}
\be{fT}
	f(T) = \chi_i T^{\beta_i}.
\ee

In terms of the parameter
\be{taudef}
	\tau(s) \equiv {2\over3}T^{-3/2}(s),
\ee
the heat flux density is
\be{Qstau}
	Q(s)=\kappa_0 a^2 \tau'(s)/R.
\ee
With this form for $Q(s)$, and the above form for $f(T)$,
we can rewrite \eqnref{Qtaueq} as
\be{taueqb}
	\ddd{}s(s\tau') + c s\tau^{-b} = 0,
\ee
in terms of a constant $c\equiv (3/2)^{-b} \chi (RB)^2/\kappa_1$, the
parameter $b\equiv2(\beta-2)/3$, and boundary condition $\tau'(1) =
0$.

This does not seem to be analytically solvable in general, but
for the temperature range below $10^5\units K$, we can set
$\chi=10^{-44}\units{W~m^3~K^{-2}}$ and $\beta=2$ in \eqnref{fT}.
For~$\beta=2$, we
have~$b=0$, and \eqnref{taueqb} is easily solved to give
\bea
	s\tau' &=& {c\over2} (1-s^2) 		\label{e:soln-st}
\\ 	\tau &=& \tau(\rho) + 		{c\over2} \left(\ln
{s\over\rho} 		+ {1\over2}(\rho^2-s^2)\right),
\label{e:soln-t}
\eea
with $\tau(\rho)=\tau(T_\rho)$.

\subsection{Consequences of these solutions}
We may make a number of observations about these solutions.  The first
is that, if we define $\delta=1-\rho$, we can rewrite \eqnref{soln-t}
as
\be{Rdelta}
	(R\delta B)^2 (1 + O(\delta)) =
{2\kappa_1\over\chi}(\tau(1)-\tau(\rho)),
\ee
and find that the radial thickness of the shell,~$R\delta$, is,
to~$O(\delta)$, dependent only on the magnetic field and the
temperature at the edge of the cell $\tau(1)$ (where
$\tau(1)\gg\tau(\rho)$, or
$T(1)\ll T_\rho$).  For $T(1)=10^4\units K$,
\eqnref{Rdelta} produces $R\delta B \approx 0.12$, giving a shell
thickness of only 12\units m for~$B=10^{-2}\units T$.

The total power removed {from} a single core in length~$\thickness$ is
\be{Prho}
	P_\rho = 2\pi\rho R\thickness Q(\rho).
\ee
If we have a large number of such filaments, each of radius~$R$, making up
a total cross-sectional area of~$A$, and we write~$P_A$ for the total
power input across this area, then we can rewrite \eqnref{Prho} with
the help of \eqnref{Qstau} to find
\be{PoverA}
	{P_A\over \thickness A} = \chi a^2 (1-\rho^2).
\ee
This indicates two things.  Firstly, since~$\rho$ has a minimum of
zero, there is a maximum to the power that can be removed {from} the
core by conduction into the shell
(namely~$P_A=10^{22}\units W$ for $\thickness=10^7\units m$,
$A=10^{13}\units{m^2}$ and $a=10^{23}\units{K~m^{-3}}$) and this
maximum is of similar order to the power of a large flare.  It is
difficult to know if this is a coincidence or not: noting that it
corresponds to the central core shrinking to a line, it seems likely
that it merely reflects the observation that there is an upper limit
to the power a given volume of plasma can radiate, and that
\eqnref{PoverA} is independent of~$n$ and~$T$ only because we have
approximated $f(T)\propto T^2$ to obtain the solution which produced it, making
the losses $n^2f(T)$ a constant.  It is nonetheless true that
$n^2f(T)$ \emph{is} approximately constant below $10^5\units K$, so
that \eqnref{PoverA} retains a qualitative general validity.

Secondly, \eqnref{PoverA} is independent of the cell radius~$R$.  This is
initially surprising, as one would expect that as the number of cells
is increased and~$R$ falls, the ratio of the conducting surface area
to the radiating volume would increase with $1/R$ and a greater power
could be removed for a given value of~$\rho$.  However,
since~$\tau'$ scales like~$R^2$, the flux density~$Q(\rho)$ itself scales
like~$R$, and so the total flux removed {from} the core scales with the same
power of~$R$ as the radiating volume.

What is happening is that the temperature gradient at the surface of the
core becomes steeper as the volume of shell it has to supply increases, and
this steepening allows us to put limits on the
permissible values of~$R$.  If the temperature gradient becomes too
great, then normal conduction will be replaced by a
more efficient anomalous mechanism which will effectively reimpose the
maximum gradient allowed by normal conduction -- the plasma will have
this gradient for some distance {from} the surface of the core, until the
solution \eqnref{soln-t} takes over, with a suitably modified inner
boundary condition, but the same outer one.  The discussion below
formally considers limits on the model's validity, but since the
maximum-gradient solution is unlikely to be stable, it describes
physical limits as well.

\begin{figure}
\ifjournalstyle
	\iffigs
		\vbox to 226pt{%
		\epsfig{file=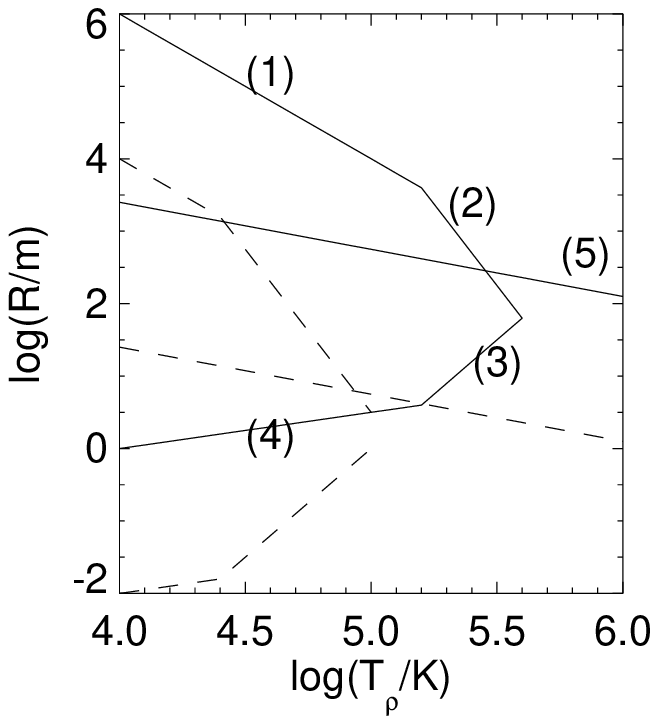}}
	\else	
		\picplace{226pt}
	\fi
\else
	\iffigs
		\includegraphics[30,0][250,226]{figr1.ps}
	\else
		\vbox to 226pt{\hrule
			\vfill\hbox{Figure 1 is in file figr1.ps}\vfill\hrule}
	\fi
\fi
\caption[]{\label{f:rplane}The restrictions on the cell radius~$R$,
obtained by comparing the temperature gradient with the ion gyroradius
and ion mean free path.  See text for explanation.  For comparison,
line~5 corresponds to the scale~$1/k_c$ in \eqnref{kc} below.  }
\end{figure}

To establish the limits on the radius in the absence of this effect,
we must insist first of all that the shell width is larger than the
ion-gyroradius,
\be{Rdeltagyro}
	R\delta \ga R_p
\ee
where
\be{Rgyro}
	R_p \approx 10^{-6} {T_\rho^{1/2}\over B},
\ee
(in the numerical expressions here and below, $B$ is in Tesla, $R$
in metres, and $n$ in \units*{m^{-3}}).  Comparison with
\eqnref{Rdelta} shows that \eqnref{Rdeltagyro}
is comfortably satisfied, independently of~$B$, for $T\la10^6\units K$.
Secondly we can define a length-scale for the
gradient,~$l_T$, via $\dd T/\dd r\sim T_\rho/l_T$ so that, using
the solution in \eqnref{soln-st} and the definition of~$\tau$ in
\eqnref{taudef}, we have
\be{lTdef}
	l_T \sim {2\times10^4 \rho \over T_\rho^{3/2} R B^2},
\ee
and the condition $l_T \ga R_p$ results in
\be{Rgrad}
	R \la 2\times10^{10} {\rho\over B T_\rho^2}.
\ee
A third length scale is provided by the ion mean free path
\be{lmfp}
	R_f \approx 2\times10^8
		\left[ {T_\rho^2\over n} = {T_\rho^3\over a}\right],
\ee
independently of the magnetic field.  Imposing $l_T \ga R_f$ results in
\be{Rmfp}
	R \la 10^{19} {\rho\over B^2 T_\rho^{9/2}}.
\ee
Together, these various constraints require that the radius~$R$ lies
between the solid lines~(1) to~(4) in \figref{rplane}.
Line~1 corresponds to \eqnref{Rgrad}, line~2 to \eqnref{Rmfp},
and lines~3 and~4 to the constraints $R\ga R_p$ and $R\ga R_f$.
All these are for the very weak field $B=10^{-4}\units T$; the dashed
lines show the corresponding constraints for $B=10^{-2}\units T$.

\begin{figure}
\ifjournalstyle
	\iffigs
		\vbox to 160mm{%
		\epsfig{file=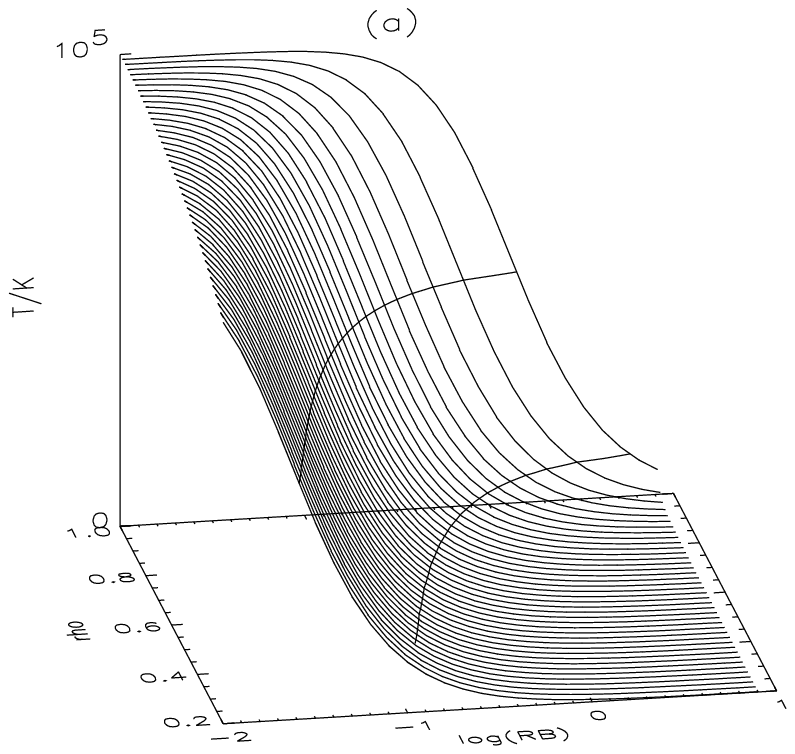}
		\epsfig{file=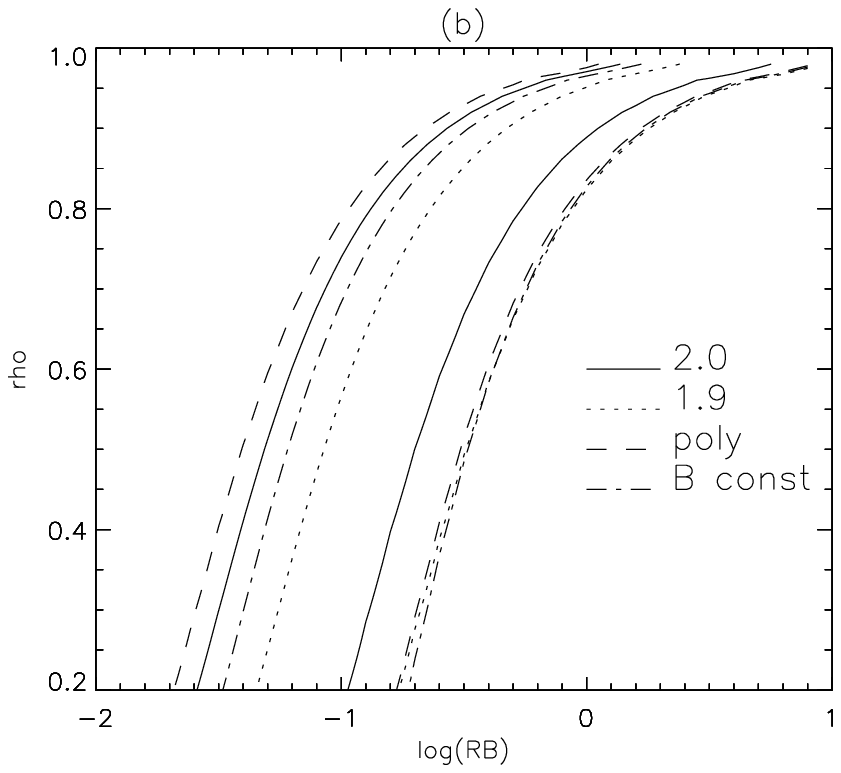}}
	\else	
		\picplace{160mm}
	\fi
\else
	\iffigs
		\includegraphics[30,0][250,226]{figtaua.ps}
		\includegraphics[30,0][250,226]{figtaub.ps}
	\else
		\vbox to 2in{\hrule
			\vfill\hbox{Figure 2a and b are in files
			figtaua.ps and figtaub.ps}\vfill\hrule}
	\fi
\fi
\caption[]{\label{f:tau}The values of $T(1)$ which are consistent with a
boundary temperature $T(\rho)=10^5\units K$, for various values of the
parameters $\rho$ and $RB$, taken {from} \eqnref{soln-t} ($\beta=2$).
The contours shown on the surface are at
$10^4\units K$ and $5\times10^4\units K$.  \figref{tau}b
shows the same two contours for $\beta=2$ as well as the
numerical solutions of the cases $\beta=1.9$ (dotted line) and the
polynomial fit (dashed line).  Finally, the dot-dash line shows the
contours for the case $B=\hbox{constant}$, discussed in
Sect.~\ref{s:lowbeta} below.}
\end{figure}

Finally, we may use the solution in \eqnref{soln-t} to find the
temperatures at $s=1$
which are consistent with $T(\rho)=T_\rho=10^5\units K$.  These are
displayed in \figref{tau}.  As can be seen there,
constraining $T(1)$ to lie
between $10^4$ and $5\times10^4\units K$ puts
a tight constraint on the possible values of $(\rho,RB)$.
That is, only a narrow range of these parameters allow solutions with a
cool outer shell in this $T$~range.

\subsection{Anomalous enhancements to $\kappa_\pe$}
\label{s:turb}
The length scales in this model are consistently very small,
due to the extreme smallness of the perpendicular
conduction coefficient~$\kappa_1$ in \eqnref{kpdef}.  Observations in
tokamaks consistently show thermal (and mass) transport coefficients
enhanced by one or two orders of magnitude over the classical values.
This is generally attributed to microturbulence (small-scale,
low-frequency, and more likely in the
magnetic than the electric field), or to some other
small-scale quasi-diffusive process, though theoretical explanations
are inconclusive
\cite{liewer85}.  Though we should, of course, be wary of exporting these
observations {from} the laboratory to the more tenuous and cooler
atmosphere of the sun, it seems safe to assume
that the scales derived here are substantial underestimates.
Effective enhancement of~$\kappa_1$ will proportionately enhance the
right-hand sides of Eqns.~(\ref{e:Rgrad}), (\ref{e:Rmfp})
and~(\ref{e:Rdelta}) and raise the
corresponding upper boundaries in \figref{rplane}.
In the Sun, ion-sound turbulence can result in a collision time much
smaller than that given above \cite{priest84}, but this occurs only
when the field has scales below $\sim 10^{21} (B/\units*T)
(n_e/\units*{m^{-3}})^{-1}(T_e/\units*K)^{-1/2}\units m$, comparable with the
lower limits given in
\figref{rplane}\draftnote{See also Hood, Priest, van der Linden and
Goossens, \dots. Or Manheimer and Cook [56)?}.

If we take a coronal magnetic field of $10^{-2}\units T$ and a boundary
temperature of $T_\rho=10^5\units K$, then \figref{rplane} and the above
two-order enhancement in the length scales give a filament scale of
$10^2\units m$, and hence require of the order of $10^9$ filaments to
fill up a flare area of $10^{13}\units{m^2}$.

\subsection{Numerical solution of the temperature profile}
\label{s:numerical}
For comparison with the above analytic results, and to confirm that
our analytic solution is not somehow pathological,
we have also solved \eqnref{Qtaueq} numerically, with
$f(T)$ represented by power-laws with various values of~$\beta$, and by a
polynomial fit \cite{keddie70} to the Cox-Tucker curve.  The solutions are
dependent on~$\beta$, but not unduly so, so that we can be confident
that our results are qualitatively insensitive to the
precise form of the loss-function.  See \figref{tau}.

\subsection{Comparison with the longitudinal temperature profile}
\label{s:comparison}

In section~\ref{s:model} above, we stated that we can ignore the
effects of longitudinal conduction, and we can now check that this is
indeed true.  We can avoid solving the full 2-d problem in the $(s,z)$
plane, and instead calculate what the longitudinal temperature
gradient would have to be for this assumption to fail.  Writing the
longitudinal flux density~$Q_\pa(z)$ as
\be{Qpa}
	Q_\pa(z) = \kappa_\pa^0 T^{5/2} T_z,
\ee
where the coefficient $\kappa_\pa^0=(2.23\times10^{-9}
\units{W~m^{-1}~K^{-7/2}})/\ln\Lambda$ \cite{spitzer62},
and~$T_z\equiv\dd T/\dd z$,
we can use the solution in \eqnref{soln-st}
to write the ratio of transverse and longitudinal flux densities as
\be{Qpepa}
	{Q_\pe\over Q_\pa} = {\chi a^2 R\over 2\kappa_\pa^0}
		\left({1\over s}-s\right) \left({3\over2}\tau(s)\right)^{5/3}
		{1\over T_z},
\ee
where~$\tau(s)$ is given by \eqnref{soln-t}.

\begin{figure}
\ifjournalstyle
	\iffigs
		\vbox to 226pt{\epsfig{file=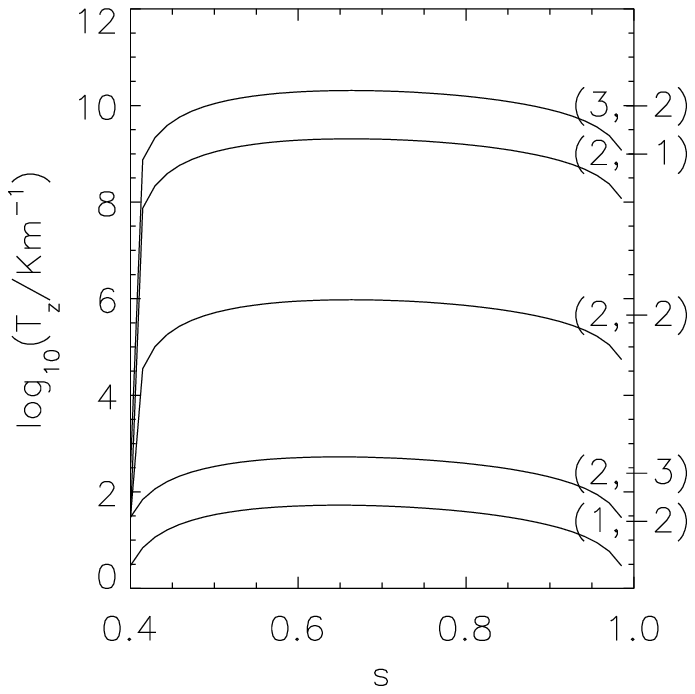}}
	\else	
		\picplace{226pt}
	\fi
\else
	\iffigs
		\includegraphics{w7173.ps}
	\else
		\vbox to 226pt{\hrule
			\vfill\hbox{Figure 1 is in file w7173.ps}\vfill\hrule}
	\fi
\fi
\caption[]{\label{f:tz}Values of~$\tilde T_z(s)/\units*{K~m^{-1}}$, the
values of the longitudinal temperature gradient~$T_z(s)$ which make the
transverse and longitudinal heat flux
densities equal, as obtained {from} \eqnref{Qpepa}.  The curves are
labelled by $(\log_{10}(R/\units*m),\log_{10}(B/\units*T))$.}
\end{figure}

In \figref{tz},
we show~$\tilde T_z(s)$, defined so that $Q_\pe>Q_\pa$ for
$T_z<\tilde T_z$, for a variety of radii and magnetic field strengths.
Note that~$\tilde T_z$ decreases towards the surface of the core
$s=\rho$: the increase in temperature here enhances the longitudinal
conduction coefficient as it suppresses the transverse one.  Note also
that~$\tilde T_z$ increases with increasing~$R$: this is initially
surprising, as one would expect the assumption $Q_\pe\gg Q_\pa$ to be
more comfortably satisfied with finer filamentation; instead this
reflects the steepening of the transverse gradient with larger~$R$,
which was noted above \eqnref{Rdeltagyro}.  In all cases, $\tilde
T_z(s=1)=0$, reflecting the boundary condition on \eqnref{taueqb} of
$\dd T/\dd s|_{s=1} = 0$, and we should also take $T_z(s=\rho) = 0$
since, by hypothesis, the temperature at the surface of the core is a
constant (\eqnref{Trho}).  It is difficult to estimate~$T_z$ in a
model-independent way, but if we take $T_z=(10^9\units*K)/\thickness
\sim 1\units{K~m^{-1}}$, we can see that for this geometry, the
transverse flux density should dominate over the longitudinal one
everywhere except at the edge of the shell,~$s=1$.

\section{Stability}
\label{s:stability}
We have considered only a static solution to the energy-balance problem.
To determine the stability of these solutions, we must examine the
heat equation
\be{heat}
	{1\over\gamma-1}\ddd pt - {\gamma\over\gamma-1}{p\over n}\ddd nt
		+ K(n,T) + \heat(n,T) = 0.
\ee
The flux term $K(n,T)$ is
\begin{eqnarray*}
	K(n,T) &=& - \nabla\cdot(\kappa_\pe\nabla T) \\
	&=& -{1\over r} \kappa_0 \ddd{}r\left[rn^2T^{-1/2} \ddd Tr\right],
\end{eqnarray*}
and the radiative loss function $\heat(n,T) = n^2 f(T)$,
so that $rK(n,T) + r\heat(n,T) = 0$ is our \eqnref{Qtaueq}.

We combine this with the equation of state, the equation of motion,
and the mass continuity condition,
and perturb the equilibrium $T$, $\rho$ and $p$ by
terms of the form
\[
	a(r,t) = a_0 + a_1 \exp(\nu t + ikr).
\]
For simplicity, we will \emph{separately} consider the limiting cases
of isobaric ($p_1=0$) and isochoric
($\rho_1=0$) perturbations -- the former match the conditions in the
earlier part of our analysis, but the latter are more appropriate for
perturbations with timescales shorter than the sound travel time.

\subsection{Isobaric perturbations}
For isobaric perturbations, $p_1=0$, the linearised heat equation,
\eqnref{heat}, becomes
$$ 
\displaylines{\ifjournalstyle\else\quad\fi 
	-{\gamma\over\gamma-1}\nu {p_0\over n_0} n_1
	- a^2\kappa_0 T_0^{-5/2} i kT_1[1/r + ik]
	\hfill\cr
	\hfill {}
	+ \heat_T T_1 + \heat_n n_1 = 0,
	\quad
	\displaylineseqno{heat-l}
	}
$$ 
where $\heat_T$ and $\heat_n$ denote partial derivatives of the heating
function $\heat(n,T)$.  Combining this with the equation of state $a=nT$,
the linearised perturbed equation of state
\[
	{n_1\over n_0} - {T_1\over T_0} = 0,
\]
the pressure $p_0=k_Ba$ (where $k_B$ is Boltzmann's constant), setting
$f(T)=\chi T^\beta$, and
setting $\gamma=5/3$, gives the real part of the growth parameter $\nu$ as
\be{Renup}
	\mathop{\mathrm{Re}}\nu = -{2\over5}{a\over k_B}\left(
		\kappa_0 T_0^{-3/2} k^2 - (2-\beta)\chi T_0^{\beta-2}\right),
\ee
which is negative when $\beta<2$ and $k>k_c$, for a critical
wavenumber~$k_c$ defined by
\be{kc}
	k_c^2 = (2-\beta) {\chi\over\kappa_1}B^2T_0^{\beta-{1\over2}}.
\ee
Thus, solutions to \eqnref{Qtaueq} are
stable to isobaric perturbations with a scale \emph{shorter} than
$1/k_c$.  This applies also to more general forms of
the loss function $f(T)$, which can be locally approximated by a
power-law~$T^\beta$.  The dependence of $1/k_c$ on~$T$ and~$B$ is
shown in \figref{rplane}.

\subsection{Isochoric perturbations}
\label{s:isochoric}
The isochoric case $\rho_1=0$ is more appropriate for perturbations which
are rapid compared to the sound time.  The same analysis as above
produces a growth rate
\be{Renurho}
	{1\over\gamma-1} {p_0\over T_0} \mathop{\mathrm{Re}}\nu
		= - n_0^2(\beta\chi T_0^{\beta-1} + \kappa_0 T_0^{-1/2} k^2),
\ee
which is certainly negative as long as~$\beta$ is positive.

We can conclude that the only unstable perturbations are
long-wavelength isobaric ones, which do not generate
sufficiently steep temperature gradients, though these perturbations
should be stabilised when the perturbation has heated enough to
provide a sufficiently large gradient.  Note that the solution
outlined above \eqnref{Rdeltagyro}, where the temperature gradient in
the inner part of the shell would have a constant maximum value, would
be unstable also -- since any local temperature enhancement will not
result in a stabilising increase in the gradient, the temperature
increase will go unchecked.

\section{Results in low-beta plasmas}
\label{s:lowbeta}
The full pressure-balance condition is
\be{pbalanceB}
	n(s)T(s) + {B^2\over2 k_B \mu_0} = a',\quad\hbox{(constant)}
\ee
which we reduced to \eqnref{pbalance} by taking the magnetic
field to be constant across the filament.  The two terms in
\eqnref{pbalanceB} are equal (for $a=10^{23}\units{K~m^{-3}}$) when
$B\approx 1.9\times10^{-3}\units T$.  If $B$ is small compared with
this, the condition in \eqnref{pbalanceB} reduces to
\eqnref{pbalance}.  For a low-$\beta$ plasma, on the other hand, a
change in the gas
pressure can be compensated by a tiny change in the magnetic field,
and \eqnref{pbalanceB} reduces to $B=\hbox{constant}$ (we thank the
referee, Dr.\ Sch\"u\ss{}ler, for emphasising this point).

In this latter case, given that the filament length and total flux are
constant, and the field lines are frozen to the plasma, we can further
take $B/n=\hbox{constant}$, and write \eqnref{Qtaueq} as
\be{taueqbB}
	\ddd{}s(s\Btau') - {B^2R^2\chi\over\kappa_1}
		s\left({\Btau\over2}\right)^{2\beta} = 0,
\ee
where $\Btau=2\sqrt T$ (compare \eqnref{taueqb}); the numerical
solution to this is displayed in \figref{tau}.  The solutions in this
case are remarkably similar to the solutions resulting {from} the
approximation in \eqnref{pbalance}.  This is explained by noting that
the change {from} \eqnref{pbalance} to \eqnref{pbalanceB} affects
\emph{both} the radiative and conductive losses:
with constraint \eqnref{pbalance}, the radiative loss term in
\eqnref{Qtaueq} is $\sim T^{\beta-2}$, and the conduction coefficient
in \eqnref{kpdef} is $\sim T^{-5/2}$; with \eqnref{pbalanceB}, these
become $T^\beta$ and $T^{-1/2}$ respectively.  This means that as the
plasma temperature rises above the equilibrium, both radiative and
conductive cooling are more efficient for the latter case than for the
former.

Because we have explicitly considered an isochoric model here, only
the stability analysis in Sect.~\ref{s:isochoric} is relevant, and we
can see that the solution for this case is stable (\eqnref{Renurho})
for all positive~$\beta$.

\section{Conclusion}
\label{s:conclusion}

We have presented a simple model of heat transport \emph{across} the
magnetic field in a highly filamented solar flare.  This is
analytically soluble in a simple case, which numerical solution shows
to be representative.

If we assume that the various potential complications do not
qualitatively change the results, then the analysis very directly
leads to predicted maximum cell sizes (\figref{rplane}) which, though
smaller than any yet observed, are larger than the lower limits
imposed by the plasma conditions.  Current uncertainties in our
understanding of anomalous enhancements to the cross-field conduction
strongly suggest that our upper limits on the cell size are
underestimates.

The upper limits we have derived here are strongly dependent on
temperature and field strength, so that in the cooler and less
magnetic environment of prominences, we would expect this mechanism to
have a more prominent role than in flares.

Given that the mechanism described here is important in flares
of all sizes, then the dependence of~$\rho$ on~$P_A$ in
\eqnref{PoverA}, and the interdependence of~$\rho$, and~$RB$
in \figref{tau} are testable predictions.  Even if we do not believe
that this mechanism governs flare geometry as generally as this, we
must regard the scale limits as delimiting a domain within which
the effects of cross-field thermal conduction dominate over
longitudinal conduction, and cannot be ignored.

\begin{acknowledgements}
NG gratefully acknowledges the support of a PPARC grant, and JCB of an
EEC contract.
\end{acknowledgements}

\def\JAA{A\&A}
\def\JApJ{ApJ}

%

\end{document}